\documentclass[twocolumn]{aastex631}
\usepackage{amsmath,amssymb,graphicx}
\usepackage{natbib}
\newcommand{\Msun}{M_\odot}
\newcommand{\MBH}{M_{\rm BH}}
\newcommand{\Mstar}{M_*}
\newcommand{\mdot}{\dot m}
\newcommand{\snrthr}{{\rm SNR}_{\rm thr}}

\shorttitle{Can LISA see inside the Little Red Dots?}

\makeatletter
\renewcommand{\frontmatter@title@above}{}
\makeatother

\begin{document}
	
	\title{Can LISA See Inside the Little Red Dots?}
	
	\author{Samuel G\'omez G\'omez}
	\email{samuel.gomez@uib.cat}
	\affiliation{Departament de F\'isica, Universitat de les Illes Balears, Palma de Mallorca, E-07122, Spain}
	\affiliation{Institute of Applied Computing \& Community Code (IAC3),
		Universitat de les Illes Balears, Palma de Mallorca, E-07122, Spain}
	
	\author{Xisco Jimenez Forteza}
	\affiliation{Departament de F\'isica, Universitat de les Illes Balears, Palma de Mallorca, E-07122, Spain}
	\affiliation{Institute of Applied Computing \& Community Code (IAC3),
		Universitat de les Illes Balears, Palma de Mallorca, E-07122, Spain}
	
	\author{Carlos Palenzuela}
	\affiliation{Departament de F\'isica, Universitat de les Illes Balears, Palma de Mallorca, E-07122, Spain}
	\affiliation{Institute of Applied Computing \& Community Code (IAC3),
		Universitat de les Illes Balears, Palma de Mallorca, E-07122, Spain}
	
	\author{Rosalba Perna}
	\affiliation{Department of Physics and Astronomy, Stony Brook University, Stony Brook, NY 11794-3800, US}
	
	\begin{abstract}
		The Little Red Dots (LRDs) are compact, broad-line sources that appear to contain accreting black holes (BHs) embedded in dense gas. Proposed models for their central environments (including quasi-stars, radiative BH envelopes and super-Eddington accretion discs) predict gas densities at the radii probed by an extreme-mass-ratio inspiral that differ by many orders of magnitude. We compute the inspiral of a $10$--$30\Msun$ BH for a range of LRD interior models, including environmental effects from the gas, and test with a mismatch criterion whether the resulting waveforms can be distinguished from vacuum and from one  another. Dense interiors (Bondi-like free-fall cavities and radiative envelopes) produce measurable deviations from vacuum in essentially every detectable event. Convective quasi-star interiors are measurable for $\MBH\lesssim10^5\Msun$ and $z\lesssim0.6$, while a co-rotating disc leaves no measurable imprint. In favorable cases the models are also mutually distinguishable. None of the 14 JWST LRDs at $z>3.9$ would be detectable with a $10\Msun$ companion, but 30 of 31 LRDs at $z<0.9$ would be if their black holes are $\sim100$ times less massive than their virial estimates, and in many of them the interior's imprint would be measurable. Across the LRD population to $z=6$, a 4-yr LISA mission would yield $0.08$--$0.8$ detectable events for $10^5\Msun$ hosts and $10\Msun$ companions ($1.5$--$15$ for $30\Msun$) if LRD nuclei host inspirals at the rate of ordinary galactic nuclei. In dense clusters, where captures are plunges, this falls to $\lesssim10^{-4}$.
	\end{abstract}
	
	
	\section{Introduction}\label{sec:intro}
	
	JWST has revealed a population of compact, red sources with broad Balmer lines at $z\sim4$--$9$, known as the ``Little Red Dots'' (LRDs) \citep{Matthee24,Greene24,Kokorev24,Kocevski26}. Their spectra combine a blue UV continuum with a red optical continuum, show little hot dust \citep{Setton25} and weak X-ray emission \citep{Maiolino25}, and often exhibit strong Balmer breaks and absorption features that require very dense gas close to the nucleus \citep{deGraaff25,Naidu25}. Their comoving density at $z\gtrsim4$ is $\sim10^{-5}$--$10^{-4}$~cMpc$^{-3}$ \citep{Kokorev24,Kocevski26}, and analogous sources have now been identified down to $z\simeq0.1$ \citep{Lin25,Lin26,Park26,Ding26}.
	
	The nature of the central engine in an LRD remains debated, although most interpretations invoke an accreting black hole (BH; e.g. \citealt{deGraaff25,Kido25,Inayoshi2025, Naidu25, BegelmanDexter26,Santarelli2026,RomanGarza2026,Curtis2026, Madau2026,Liu2026,Chon2026,Torralba2026,
		Liu2026b,Kaaz26,Zhou26}). The broad Balmer lines, much wider than the forbidden lines, could arise from compact, unresolved regions, as in broad-line active galactic nuclei \citep{Matthee24,Greene24,Kocevski26}, while the Balmer absorption and breaks require dense gas close to a luminous central source \citep{deGraaff25,Naidu25,Rusakov25}. The mass of the central BH, if present, is uncertain. If the broad lines are virialized, the inferred masses are $\MBH\sim10^6$--$10^8\,\Msun$, making them overmassive relative to their hosts. If the line widths instead arise from electron scattering in a dense cocoon, the masses are $\sim100$ times lower \citep{Rusakov25}. The surrounding gas could form the envelope of a quasi-star (QS), in which a black hole grows inside a radiation-supported star \citep{CoughlinBegelman24,BegelmanDexter26,Hassan26}; a radiative envelope (RE) surrounding an Eddington-limited BH \citep{Kido25}; or a super-Eddington accretion disc viewed at high inclination \citep{Zhou26}, or its associated wind \citep{Kaaz26}. Although these models are designed to reproduce similar spectra, they predict gas densities that differ by many orders of magnitude at $r\sim10^{10}$--$10^{13}$~cm.
	
	This difference is directly relevant for LISA: these are the radii probed during the final years of an extreme-mass-ratio inspiral (EMRI) of a stellar-mass compact object into a $10^4$--$10^6\,\Msun$ BH \citep{AmaroSeoane17,Babak17}. Environmental effects from the surrounding gas modify the inspiral, and the accumulated phase shift can be measurable \citep{Kocsis11,Barausse14,Speri23}. How frequently such inspirals occur in LRDs remains uncertain. Ordinary galactic nuclei host $\sim10^{-7}$--$10^{-6}$ EMRIs per year \citep{Babak17}. \citet{KritosSilk26a,KritosSilk26b} model LRD nuclei as dense, mass-segregated clusters that feed the central BH through the loss cone, but predominantly through direct plunges rather than multi-year inspirals.
	
	The large differences in gas density predicted by these central-environment models therefore offer a potential way to distinguish them with LISA EMRIs. This Letter investigates whether an EMRI in an LRD can discriminate among the broad classes of models described in Sec.~\ref{sec:models}. The  methods used to compute dephasing during an inspiral are detailed in Sec.~\ref{sec:method}, while results are presented in Secs.~\ref{sec:scan}--\ref{sec:rates}. We adopt a flat $\Lambda$CDM cosmology with $H_0\simeq67.7$~km\,s$^{-1}$\,Mpc$^{-1}$ and $\Omega_m\simeq0.31$.
	
	\section{Interior models}\label{sec:models}
	
\begin{figure}[t]
		\centering
		\includegraphics[width=\columnwidth]{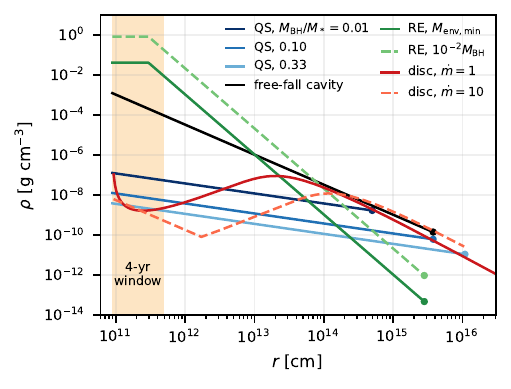}
		\caption{Gas density profiles of the interior models at $\MBH=10^5 \Msun$. Blue: quasi-star (QS) for three mass ratios $\MBH/\Mstar$. Black: free-fall cavity. Green: radiative envelope (RE) at the two ends of the envelope-mass range considered, the minimum bound mass (solid) and $10^{-2}\MBH$ (dashed).
			Red: $\alpha$-disc midplane density for $\mdot=1$ (solid) and $10$ (dashed). Spherical models are shown out to the envelope base $R_i$ or the photosphere (dots). The shaded band marks the orbital radii during the final 4~yr before the ISCO for a $10\,\Msun$ companion.
		}
		\label{fig:profiles}
	\end{figure}

	Figure~\ref{fig:profiles} shows the density profiles of three proposed models for the close environment of a BH of $\MBH=10^5 \Msun$, together with the standard Bondi-like solution as a reference. Over the radii probed during the final four years of inspiral in the LISA band, the densities span nine orders of magnitude.
	
	\paragraph{Quasi-star (QS)} These systems consist of a BH growing inside a massive, radiation-supported envelope of mass $\Mstar$ \citep{Begelman06,Begelman08}. 
	The quasi-star radiates at the Eddington luminosity of its envelope, $L=L_{\rm Edd}(\Mstar)=4\pi G\Mstar c/\kappa$ ($\kappa=0.34$~cm$^2$\,g$^{-1}$ is the electron-scattering opacity), which near the BH is carried outward by vigorous (``saturated'') convection with efficiency $\eta$. Requiring convection to carry $L$ at every radius, together with hydrostatic balance, gives the universal interior \citep{CoughlinBegelman24,BegelmanDexter26},
	\begin{equation}
		\rho_{\rm QS}(r)=\frac{9}{16\sqrt2\,\pi}\,\frac{L}{\eta\,(G\MBH)^{3/2}}\,r^{-1/2},
		\label{eq:QS}
	\end{equation}
	with $\eta=0.1$. 
	The sound speed, $c_s^2=8G\MBH/9r$, makes the companion's orbit transonic, $v/c_s\simeq1.06$. 
	In the evolutionary models of \citet{Hassan26}, which reach $\MBH/\Mstar=0.33$, approximately $70\%$ of the lifetime is spent at $\MBH/\Mstar={0.1-0.33}$. In an exploratory model with eruptive winds, however, the envelope is consumed early and the bright phase is concentrated at $\MBH/\Mstar\lesssim0.02$.
	We therefore take $\MBH/\Mstar=0.1$, typical of the no-wind evolution, as the fiducial value, and bracket it with the gas-rich early stage,  $\MBH/\Mstar=0.01$ (envelope $\Mstar=100\,\MBH$), 
	and the gas-poor late stage, $\MBH/\Mstar=0.33$ (envelope
	$\Mstar=3\,\MBH$), the latest stage evolved by Hassan et al.
	
	The convective efficiency $\eta$ is poorly constrained, while convective eddies moving at a sizeable fraction of the orbital speed make the gas-at-rest drag prescription approximate; together, these effects make the drag uncertain by a factor of several. Our quasi-star results should therefore be regarded as upper limits.
	
	\paragraph{Bondi-like free-fall cavity} As an extreme dense case we take the profile of gas falling freely onto the BH inside its Bondi-like radius
	$R_i$, $\rho_{\rm B}=\rho_i(r/R_i)^{-3/2}$. We anchor it at the boundary values of
	\citet{Hassan26}, $R_i=1.2\times10^{15}$~cm and $\rho_i=4.5\times10^{-10}$~g\,cm$^{-3}$ for $\MBH/\Mstar=0.10$ and $\Mstar=10^5\,\Msun$, rescaled as
	$R_i\propto\Mstar^{1/2}$ and $\rho_i\propto\Mstar^{-1/2}$, with an adiabatic sound speed. At $1$--$3\,r_{\rm ISCO}$ ($r_{\rm ISCO}=6G\MBH/c^2$) it is $10^4$--$10^5$ times
	denser than the quasi-star model from Eq.~(\ref{eq:QS}). Notice that it is not a self-consistent interior: gas with this profile
	falls in at nearly the free-fall speed, comparable to the orbital speed, and at these densities would empty the cavity within decades.  Treating it as at rest
	therefore overestimates the drag. 
	
	\paragraph{Radiative envelope (RE)} Unlike the quasi-star, whose envelope outweighs the BH, \citet{Kido25} place a BH radiating at its own Eddington
	luminosity inside an envelope much lighter than itself, $M_{\rm env}\ll\MBH$. The luminosity then diffuses outward radiatively, such that the interior is given by
	\citep{Ulmer98}
	\begin{equation}
		\rho_{\rm RE}(r)=\frac{M_{\rm env}}{4\pi\Lambda r^3},\quad
		\Lambda=\ln\frac{r_{\rm ph}}{r_{\rm bnd}}-1.8,
		\label{eq:kido}
	\end{equation}
	between the inner boundary $r_{\rm bnd}=10\,R_S$, where the luminosity is injected and inside which we hold $\rho$ constant, and the photosphere $r_{\rm
		ph}\simeq9\times10^{14}(\MBH/10^4\Msun)^{1/2}$~cm. The sound speed, $c_s^2=G\MBH/3r$, sets the drag and makes the orbit supersonic, $v/c_s\simeq\sqrt3$. 
	The shape of the profile is fixed, but its normalisation is set by the envelope mass. We consider the range $M_{\rm env,min}\le M_{\rm env}\le10^{-2}\MBH$ and compute environmental effects during EMRIs for these two limiting values, where
	the minimum mass that stays bound against the injected energy sets $M_{\rm env,min}\simeq50\,(\MBH/10^5\Msun)^{3/4}\,\Msun$ \citep{Kido25}.
	The model is tuned to $\MBH\sim10^7\,\Msun$, i.e.\ to the virial masses; around lighter BHs such an envelope would be too faint to appear as an LRD,  so our results there illustrate this class of models rather than a realistic LRD.  
	
	\paragraph{Super-Eddington disc} We use the $\alpha$-disc of \citet{Shakura73}, obtaining the midplane density at each radius by solving the local structure equations, so that its three classical zones --dominated by radiation pressure with electron scattering, gas pressure with electron scattering, and gas pressure with free-free opacity-- join smoothly.
	We assume a zero-torque inner edge, $\alpha=0.1$, and truncation where the disc becomes self-gravitating; inside the radius where the thin solution reaches
	$H/r=1$, which exists for $\mdot\equiv\dot M/\dot M_{\rm Edd}>2.7$ (with $\dot M_{\rm Edd}=L_{\rm Edd}/(\epsilon c^2)$, $\epsilon=0.1$), we use a slim, advective
	disc. We consider two accretion rates: $\mdot=1$, an Eddington-limited disc that stays thin down to the ISCO, and $\mdot=10$, a super-Eddington disc with a
	slim inner zone. The published LRD discs have $\mdot\sim10^2$--$10^3$ \citep{Zhou26,Kaaz26}, but mass loss limits the inflow near the BH to $\mdot\sim10$
	\citep{Zhou26}, so in the LISA window their density is close to our $\mdot=10$ case.
	There, with the slim zone matched to the thin disc at the trapping radius, the disc is $2$--$10$ times less dense than the fiducial quasi-star ($\MBH/\Mstar=0.1$).
	
	\section{Inspiral, dephasing and resolvability}\label{sec:method}
	
	We follow a circular orbit of a companion of mass $m$ at radius $r$ in the potential of the black hole and the enclosed gas, $M_t=\MBH+m+M_{\rm gas}(<r)$, with $\Omega=(GM_t/r^3)^{1/2}$ and $v=\Omega r$. 
	In the spherical models, the companion grows by Bondi--Hoyle--Lyttleton accretion \citep{HoyleLyttleton39,BondiHoyle44}, $dm/dt=\dot m_{\rm BHL}=4\pi G^2m^2\rho/(v^2+c_s^2)^{3/2}$. Since the accreted gas carries no angular momentum, the orbital angular momentum, $m\sqrt{GM_tr}$, changes through both the torques and the growth of $m$. We integrate
	\begin{align}
		& \frac{m v}{2}\left(1+\frac{r}{M_t} \frac{dM_{\rm gas}}{dr}\right)\dot r =
		-\frac{\dot E_{\rm GW}}{\Omega}-F_{\rm DF}\,r-\dot m_{\rm BHL}\,rv,
		\nonumber\\
		&\frac{\dot f}{f}=\frac12\left(\frac{r}{M_t}\frac{dM_{\rm gas}}{dr}-3\right)\frac{\dot r}{r}
		+\frac{\dot m_{\rm BHL}}{2M_t},\label{eq:fdot}
	\end{align}
	with $f=\Omega/\pi$ the GW frequency. The GW luminosity is given by the quadrupole formula \citep{Peters64}, $\dot E_{\rm
		GW}=\frac{32}{5}\frac{G}{c^5}\mu^2r^4\Omega^6$, with reduced mass $\mu=m\,(\MBH+M_{\rm gas}(<r))/M_t$, and the resulting chirp rate includes post-Newtonian
	corrections. Gaseous dynamical friction follows  $F_{\rm DF}=4\pi G^2m^2\rho\,I/v^2$ \citep{Ostriker99},
	where the Coulomb factor $I$ takes Ostriker's subsonic or supersonic form depending on the Mach number $v/c_s$, with the wake size limited to the orbital
	radius \citep{KimKim07}.  In the spherical
	models the gas is taken to be at rest, so the speed relative to the gas is the orbital speed $v$.  For the co-rotating disc, we set $F_{\rm DF}=\dot m_{\rm BHL}=0$ and retain only the enclosed gas mass, $M_{\rm gas}(<r)$, computed from the disc surface density.
	
	To calculate the dephasing, we integrate two inspirals of the same system, one including the environmental effects of the gas (chirp rate $\dot f_{\rm env}$) and one in vacuum ($\dot f_{\rm vac}$), from $f_0$, the frequency at which the vacuum chirp reaches the ISCO after $T_{\rm obs}/(1+z)$ in the source frame, to $f_{\rm ISCO}$, using $T_{\rm obs}=4$~yr. Aligning the two at the ISCO, the accumulated difference in GW phase is
	\begin{equation}
		\Delta\phi=2\pi\int_{f_0}^{f_{\rm ISCO}}f\left(\frac{1}{\dot f_{\rm env}}-
		\frac{1}{\dot f_{\rm vac}}\right)df .
		\label{eq:dephasing}
	\end{equation}
	
	We call a model \emph{perturbative} when the gas mass inside the initial orbit is below the companion mass $m$ and the total environmental correction to the chirp rate (drag, accretion, and enclosed mass) remains below the GW-driven rate, $\epsilon_{\max}=\max(\dot f_{\rm env}/\dot f_{\rm vac}-1)<1$. Outside this regime, the inspiral is gas-dominated and a vacuum template is no longer a sensible reference.
	
	The sky-averaged SNR over the last $T_{\rm obs}$ observed years is
	\begin{equation}
		{\rm SNR}^2=4\int_{f_{\min}}^{f_{\max}}\frac{|\tilde h(f)|^2}{S_n(f)}\,df,
		\label{eq:snr}
	\end{equation}
	where $\tilde h(f)$ is the Fourier-domain GW strain and $f_{\min}=f_0/(1+z)$ and $f_{\max}=f_{\rm ISCO}/(1+z)$ are the observed frequencies at the start and end of the last $T_{\rm obs}$. We use the leading-order amplitude with the redshifted chirp mass and luminosity distance, and the LISA sensitivity of
	\citet{Robson19} including Galactic confusion noise. An event is \textit{detectable} if SNR$>10$; for a $10\,\Msun$ companion this horizon reaches
	$z\simeq2.2$ at $\MBH=10^5\,\Msun$ and shrinks for lighter and heavier hosts, and a $30\,\Msun$ companion extends it to $z\simeq5$.
	
	A dephasing of $\sim0.1$~rad is often used as the threshold for a measurable effect~\citep{Maselli:2022}. However, part of a smooth environmental phase can
	be absorbed into the coalescence time $t_c$ and phase $\phi_c$ of a vacuum template, leaving a smaller residual, so a larger SNR is needed than this
	threshold suggests. We therefore use the frequency-domain phase difference in the stationary-phase approximation, $\Delta\Psi(f)=2\pi f\Delta t(f)-\Delta\phi(f)$, where $\Delta t(f)$ is the time difference between the two inspirals at frequency $f$, and compute the mismatch $\mathcal{M}=1-\max_{t_c,\phi_c}\langle h_{\rm env}|h_{\rm vac}\rangle$ from the normalised overlap weighted by $|\tilde h|^2/S_n$. The environment might be
	resolvable only when
	\begin{equation}
		{\rm SNR}>\snrthr\equiv\frac{2}{\sqrt{2\,\mathcal{M}}},
		\label{eq:thr}
	\end{equation}
	which is the distinguishability criterion of \citet{Lindblom08} made a factor of two stricter, $\|\delta h\|={\rm SNR}\sqrt{2\mathcal M}>2$. We evaluate it
	as a function of the BH mass and redshift, for a companion of $m=10$ or $30\,\Msun$. We count an event as \textit{resolvable} if it is detectable (SNR$>10$),
	satisfies Eq.~(\ref{eq:thr}), and is perturbative. Notice that fitting also the masses could absorb more of the environmental phase, so our $\snrthr$ are lower limits.
	For a fully dephased signal, $\mathcal{M}\to1$ and $\snrthr\to\sqrt2$, below the detection threshold.
	
	\section{Dephasing across mass ratio and redshift}\label{sec:scan}
	
\begin{figure*}[t]
		\centering
		\includegraphics[width=\textwidth]{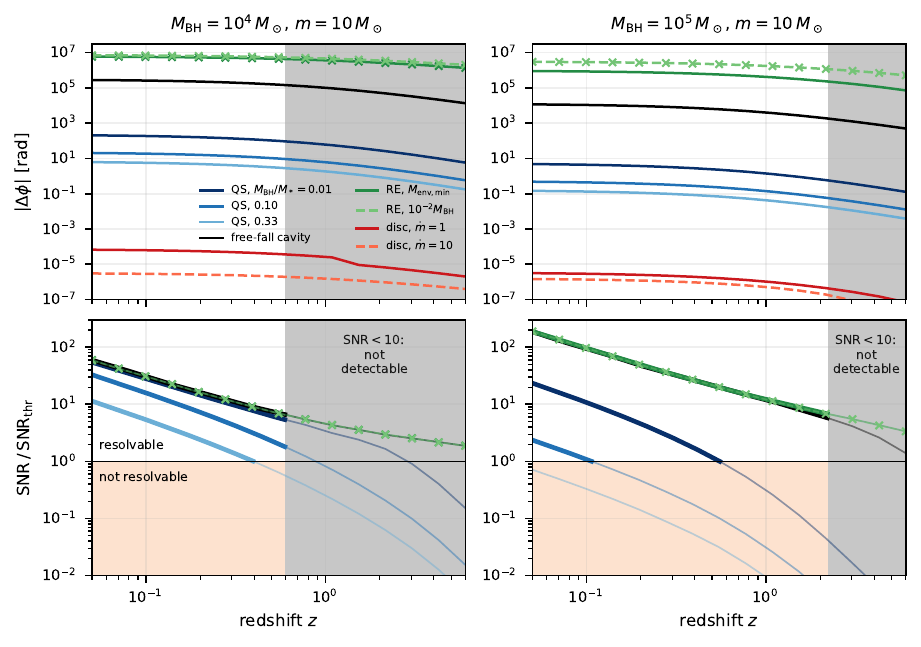}
		\caption{Top: environmental dephasing of the last 4 observed years versus redshift for
			$\MBH=10^4$ (left) and $10^5\,\Msun$ (right), with a $10\,\Msun$ companion. The disc dephasing is so small that it is computed directly from the integral in Eq.~(\ref{eq:dephasing}) rather than by integrating the orbit, whose numerical precision is insufficient at this level.
			Bottom: SNR$/\snrthr$ for the spherical
			models. An interior is resolvable above the black line (below it, light shading), outside
			the grey SNR$<10$ band and where perturbative (bold); crosses mark non-perturbative cases, in which the gas drives the inspiral and the dephasing saturates.}
		\label{fig:zscan}
	\end{figure*}

	We now apply these criteria to each interior model. Figure~\ref{fig:zscan} shows, for $\MBH=10^4$ and $10^5\,\Msun$ and a $m=10\,\Msun$ companion, how the
	dephasing and the ratio SNR$/\snrthr$ change with redshift; an interior is resolvable wherever this ratio exceeds unity in a detectable, perturbative event.
	
	For quasi-stars, the dephasing nearby ($z=0.05$) ranges from $\Delta\phi\simeq0.15$~rad ($\MBH=10^5\,\Msun$, $\MBH/\Mstar=0.33$) to $\simeq200$~rad ($\MBH=10^4\,\Msun$, $\MBH/\Mstar=0.01$). It is larger for lighter hosts and for earlier, gas-richer stages, since
	the density scales as $(\MBH/\Mstar)^{-1}$, and it decreases with redshift as the source-frame window shrinks. Although these values exceed the $\sim0.1$~rad
	rule, the mismatch criterion requires a much higher SNR (Sec.~\ref{sec:method}), so the imprint is resolvable only nearby: at $\MBH=10^4\,\Msun$ up to the
	detection horizon ($z\simeq0.6$) for $\MBH/\Mstar\le0.1$; at $10^5\,\Msun$ only to $z\simeq0.55$ at the gas-rich end ($\MBH/\Mstar=0.01$), to $z\simeq0.1$ at the fiducial
	$\MBH/\Mstar=0.1$, and not at all at the gas-poor end ($\MBH/\Mstar=0.33$).
	
	The denser interiors dephase by $\Delta\phi \sim 10^2$--$10^6$~rad where perturbative, so strongly that the environment-dressed and vacuum waveforms are
	completely different: the mismatch approaches unity and $\snrthr$ drops to $\simeq1.5$, well below the detection threshold. Whenever the inspiral remains
	perturbative, these interiors are therefore resolvable in every detectable event, out to the detection horizon. For the free-fall cavity this holds at both
	masses, so beyond $z\simeq0.6$ at $\MBH=10^5\,\Msun$ it remains resolvable where the quasi-star is not. The radiative envelope is more restrictive, because its
	steep $r^{-3}$ profile piles the gas up close to the BH. It is perturbative only for envelopes close to the minimum bound mass and for the heavier host
	($\MBH=10^5\,\Msun$), where it is resolvable out to the horizon. For lighter hosts, or for $M_{\rm env}=10^{-2}\MBH$, the gas inside the orbit outweighs the
	companion or the drag exceeds GW emission. 
	The inspiral is then gas-driven, and its dephasing saturates at the total number of vacuum cycles in the window, $\sim10^6$--$10^7$~rad,  independently of the density, which is why the two envelope curves nearly coincide at $\MBH=10^4\,\Msun$ in Fig.~\ref{fig:zscan}. Such an imprint requires a non-perturbative treatment that is beyond the scope of this work.
	
	The disc behaves very differently. Because its gas co-rotates with the companion, there is no dynamical friction or accretion drag, and the only effect
	we include is the enclosed disc mass, which slightly modifies Kepler's law. A thin disc holds very little mass inside the companion's orbit, at most $\sim3\times10^{-7}\,\Msun$ at the start of the 4-yr window, so the dephasing is just $\Delta\phi \sim 10^{-7}$--$10^{-5}$~rad. This is below the numerical floor of our mismatch
	calculation, so we can only bound $\snrthr\gtrsim3\times10^3$, far above any achievable SNR. The disc therefore leaves no resolvable imprint through its
	enclosed mass. Migration torques, which require a dedicated treatment (thin versus slim disc regimes, gap opening), are not modelled; a rough Type-I estimate suggests they could matter only for thin discs around the lightest hosts \citep{Kocsis11,Speri23}.
	
	We can apply the same mismatch criterion to two waveforms including the environmental effects of different LRD models, rather than comparing one with vacuum.
	The free-fall cavity and radiative envelope can be distinguished from every other model wherever they are detectable and perturbative. A disc is indistinguishable from vacuum, so distinguishing it from a quasi-star is equivalent to resolving the quasi-star imprint. Different quasi-star stages can be distinguished only nearby: at $\MBH=10^5\,\Msun$, the gas-rich stage ($\MBH/\Mstar=0.01$) separates from the later stages and from a disc to $z\simeq0.5$, while $\MBH/\Mstar=0.1$
	separates from $\MBH/\Mstar=0.33$ only to $z\simeq0.08$, and the gas-poor stage is never distinguishable from a disc. These limits are however optimistic, since the convective
	efficiency $\eta$ is degenerate with the stage. 
	
	\section{Application to the observed LRDs}\label{sec:samples}
	
\begin{figure}[t]
		\centering
		\includegraphics[width=\columnwidth]{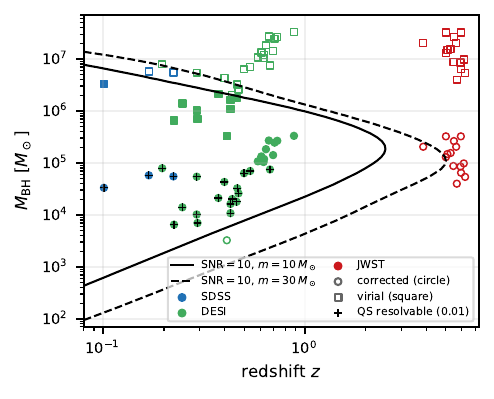}
		\caption{The observed LRDs in the $(z,\MBH)$ plane: SDSS (blue), DESI (green) and JWST (red), at the scattering-corrected (circles) and virial (squares) masses.
			Filled symbols are detectable (SNR$>10$) with a companion of $m=10\,\Msun$; every detectable object would also resolve a free-fall cavity, and a minimum-mass
			radiative envelope wherever it is perturbative, while the radiative envelope with $M_{\rm env}=10^{-2}\MBH$ is non-perturbative and the disc is never
			resolvable. Plus signs mark a resolvable quasi-star interior at $\MBH/\Mstar=0.01$. Black curves: SNR$=10$ horizon for $m=10\,\Msun$ (solid) and $m=30\,\Msun$ (dashed).}
		\label{fig:samples}
	\end{figure}

	We apply the previous calculation to two spectroscopic samples, shown in Fig.~\ref{fig:samples}. The first comprises 14 JWST LRDs at $z=3.9$--$6.2$ with broad H$\alpha$ virial masses from the Skyfire census of the CEERS field \citep{Kocevski26}. The second comprises 31 spectroscopically confirmed LRDs at
	$z=0.10$--$0.88$: three SDSS objects \citep{Lin25}, 27 DESI objects \citep{Lin26}, and one from \citet{Park26}. For 11 of the low-$z$ objects, the mass is
	inferred from the line width, while for the other 20 it is inferred from the bolometric luminosity, assuming the median Eddington ratio, $L_{\rm bol}/L_{\rm
		Edd}$, of the 11. For each object, we use both the virial mass $M_{\rm vir}$ and the scattering-corrected mass, $M_{\rm vir}/100$ \citep{Rusakov25}. We assume a $m=10\,\Msun$
	companion, first assess whether each signal is detectable, and then whether the imprint of each interior model is resolvable.

	For the JWST sample, none of the 14 objects is detectable. At the corrected masses, $\MBH=(4$--$30)\times10^4\,\Msun$, the median SNR at $z\simeq5$ is
	$\simeq4.9$; at the virial masses it decreases to $\sim10^{-3}$. Companions of mass $m=23$--$41\,\Msun$ would bring the most favourable to the median object
	above SNR$=10$.
	
	For the low-redshift sample, most local LRDs lie within the horizon at the corrected masses: 30 of 31 are detectable, with ${\rm SNR}=13$--$85$. Assuming every object is a quasi-star, the imprint is resolvable in 5--20 of them depending on the stage, all with $\MBH\lesssim10^5\,\Msun$ (0--26 if the envelope mass is instead set by the luminosity). The
	free-fall cavity is resolvable in all 30 detectable objects, and the minimum-mass radiative envelope ($M_{\rm env}=M_{\rm env,min}$) in the 14 where it is
	perturbative, although this model belongs with the virial masses. 
	
	At the virial masses, $\MBH=10^{5.5}$--$10^{7.5}\,\Msun$, only 10 objects are detectable,
	all with $\MBH\le10^{6.5}\,\Msun$ and $z\le0.46$; in these, the free-fall cavity and the minimum-mass radiative envelope are resolvable, while the quasi-star is not.
	The radiative envelope with $M_{\rm env}=10^{-2}\MBH$ is non-perturbative in every object (the gas inside the orbit outweighs the companion), so the inspiral
	is gas-driven and outside the scope of this analysis. The disc is not resolvable in any object under either mass hypothesis.
	
	\section{Event rates}\label{sec:rates}
	
\begin{figure}[t]
		\centering
		\includegraphics[width=\columnwidth]{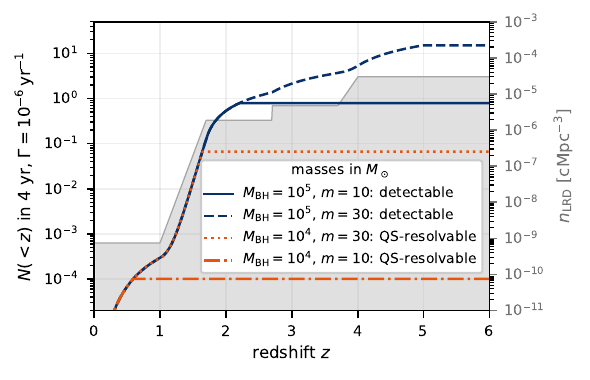}
		\caption{ Cumulative number of events in 4~yr, $N(<z)$, for multi-year inspirals at $\Gamma=10^{-6}$~yr$^{-1}$ per nucleus. Dark blue: detectable events for $\MBH=10^5\,\Msun$ with companions of $m=10\,\Msun$ (solid) and $m=30\,\Msun$ (dashed); every such event also resolves a free-fall cavity. Orange: events with a resolvable quasi-star interior at $\MBH/\Mstar=0.075$ in the most favourable host, $\MBH=10^4\,\Msun$, with $m=30\,\Msun$ (dotted) and $m=10\,\Msun$ (dash-dotted). Grey shading, right axis: the adopted LRD number density $n_{\rm LRD}$. Below their horizons all events count, so the curves coincide; each flattens at the SNR$=10$ horizon of its host and companion, $z\simeq2.2$, $5.0$, $1.6$ and $0.6$ respectively, beyond which no event is detectable.}
		\label{fig:rates}
	\end{figure}

	The prospects for observing these signals depend on how often multi-year inspirals occur in LRD nuclei. We estimate here the number that LISA could detect or resolve. For a per-nucleus event rate $\Gamma$,
	the expected number in a mission of duration $T_{\rm obs}$ is
	\begin{equation}
		N=T_{\rm obs}\int_{0}^{6} n_{\rm LRD}(z)\,\frac{\Gamma}{1+z}\,
		\chi(\MBH,z)\,\frac{dV_c}{dz}\,dz ,
		\label{eq:N}
	\end{equation}
	where $1/(1+z)$ converts the rate to the observer frame, $dV_c/dz$ is the all-sky comoving volume element, and the selection function $\chi(\MBH,z)$ is 1
	when an event at that mass and redshift meets the chosen criterion (detectable, SNR$>10$, or resolvable) and 0 otherwise. For the LRD number density we follow \citet{Lin26} and \citet{Ma25}: from $\sim7.5\times10^{-10}$~cMpc$^{-3}$ at $z<1$, it rises by more than four orders of magnitude to $z>4$ (grey shading in Fig.~\ref{fig:rates}). Since the LRD BH mass function is unknown, we evaluate $N$ assuming that every LRD has the same mass, $\MBH=10^4$, $10^5$ or $3\times10^5\,\Msun$,
	spanning the scattering-corrected mass range where LISA is most sensitive (see Table~\ref{tab:rates}).
	
	The largest uncertainty lies in the per-nucleus rate $\Gamma$. In the dense star clusters that \citet{KritosSilk26a,KritosSilk26b} propose for LRD nuclei, the central BH
	often captures stellar-mass BHs, at $\sim6\times10^{-8}$--$10^{-5}$~yr$^{-1}$ per nucleus for $\MBH=10^6\,\Msun$, but these arrive on nearly radial
	orbits and plunge directly into the BH, instead of spiralling in for years as our measurement requires. Multi-year inspirals are far rarer in
	their model, $\sim3\times10^{-12}$~yr$^{-1}$ per nucleus, which in Eq.~(\ref{eq:N}) gives $\lesssim4\times10^{-5}$ events in 4~yr even over the full
	population. We therefore adopt, as an optimistic scenario, the EMRI rate of ordinary galactic nuclei, $\Gamma=10^{-7}$--$10^{-6}$~yr$^{-1}$ \citep{Babak17},
	four to five orders of magnitude higher. Even at this rate, the 31 known low-$z$ LRDs at their corrected masses yield only $\lesssim10^{-4}$ events, while
	the full population contributes far more, because the number density and the comoving volume both grow with redshift. 
	
	\begin{table}[t]
		\caption{Expected number of EMRI events in a 4-yr LISA mission from all LRDs at $z<6$, assuming that every LRD hosts a BH of mass $\MBH$ and that the
			companion has mass $m$, at the ordinary-nucleus rate $\Gamma=10^{-6}$~yr$^{-1}$; the numbers scale linearly with $\Gamma$. "Detectable'' counts events with SNR$>10$, each of which also resolves a free-fall cavity. ``QS resolvable'' counts events in which a quasi-star interior is
			resolvable, at the median stage implied by the luminosities of the $z\sim5$ sample, $\MBH/\Mstar=0.075$, and at the gas-rich stage, $\MBH/\Mstar=0.01$.}
		\label{tab:rates}
		\centering\footnotesize
		\setlength{\tabcolsep}{4pt}
		\begin{tabular}{|c|c|c|c|c|}
\noalign{\global\arrayrulewidth=0.8pt}\hline\noalign{\global\arrayrulewidth=0.4pt}
$M_{\rm BH}$ & $m$ & detectable & \multicolumn{2}{c|}{QS resolvable}\\
\cline{4-5}
$[M_\odot]$ & $[M_\odot]$ & & $M_{\rm BH}/M_*=0.075$ & $0.01$\\
\hline
$10^4$ & 10 & $10^{-4}$ & $10^{-4}$ & $10^{-4}$\\
 & 30 & 0.07 & 0.07 & 0.07\\
\noalign{\global\arrayrulewidth=0.2pt}\hline\noalign{\global\arrayrulewidth=0.4pt}
$10^5$ & 10 & 0.79 & $2\times10^{-6}$ & $9\times10^{-5}$\\
 & 30 & 15 & $2\times10^{-5}$ & $2\times10^{-4}$\\
\noalign{\global\arrayrulewidth=0.2pt}\hline\noalign{\global\arrayrulewidth=0.4pt}
$3\times10^5$ & 10 & 0.91 & $6\times10^{-9}$ & $4\times10^{-6}$\\
 & 30 & 3.3 & $2\times10^{-7}$ & $2\times10^{-5}$\\
\noalign{\global\arrayrulewidth=0.8pt}\hline\noalign{\global\arrayrulewidth=0.4pt}
\end{tabular}

	\end{table}

	Figure~\ref{fig:rates} shows the cumulative number of LRD-hosted EMRIs expected in a four-year mission, and Table~\ref{tab:rates} summarises the totals. At the ordinary-nucleus rate, hosts with $\MBH=10^5\,\Msun$ give $N\simeq0.08$--$0.8$ detectable events with a companion of $m=10\,\Msun$ and $N\simeq1.5$--$15$ with $m=30\,\Msun$, every one of which would resolve a free-fall cavity. Because the LRD density rises so steeply, these events come from just inside the detection horizon, at $z\sim2$ for $m=10\,\Msun$ and up to $z\sim5$ for $m=30\,\Msun$, consistent with the companion masses needed to detect the JWST sample (Sec.~\ref{sec:samples}). A quasi-star imprint is much rarer. It requires light hosts at low redshift, where LRDs are $10^3$--$10^4$ times scarcer than at $z\sim2$--$5$. At the median stage implied by the $z\sim5$ luminosities, $\MBH/\Mstar=0.075$, the best case, $\MBH=10^4\,\Msun$ with $m=30\,\Msun$, gives $N\lesssim0.07$; at the gas-rich stage, $\MBH/\Mstar=0.01$, favoured if eruptive winds operate (Sec.~\ref{sec:models}), the most favourable host mass gives $N\lesssim0.4$.
	
	\section{Conclusions}\label{sec:conclusions}
	
	A LISA EMRI can in principle distinguish the proposed LRD interiors. Dense interiors---a Bondi-like free-fall cavity or a radiative envelope---are resolvable in any detected, perturbative event, while a quasi-star with a convective interior is resolvable with a $10\,\Msun$ companion only for $\MBH\lesssim10^5\,\Msun$ at
	$z\lesssim0.6$;  a disc, on the other hand,  is not resolvable through its enclosed mass. The JWST LRDs are beyond reach with a $10\,\Msun$ companion, but at the
	scattering-corrected masses the local ones lie within the horizon. The limitation is the event rate: the known local objects yield $\lesssim10^{-4}$ inspirals in four years, while the full population to $z=6$ gives $0.08$--$0.8$ ($m=10\,\Msun$) or $1.5$--$15$ ($m=30\,\Msun$) events for $\MBH=10^5\,\Msun$ hosts if LRD nuclei host inspirals at the ordinary-nucleus rate.
	
	A single such event would be valuable well beyond these models. On its own, the waveform measures the redshifted BH and companion masses with high precision, the BH spin, the luminosity distance and an approximate sky position, and, if the gas imprint is resolvable, the gas density and its slope within a few tens of Schwarzschild radii. Combined with electromagnetic observations, the distance and sky
	position can identify the host, whose spectroscopic redshift yields the source-frame mass. Comparing that mass with the virial estimate would test directly whether the broad lines are virial or broadened by electron scattering, and, together with the observed luminosity, may further provide an indication on the quasi-star stage, $\MBH/\Mstar$.
	The GW-inferred density close to the BH and the dense gas inferred from Balmer absorption further out would together constrain the gas distribution over several orders of magnitude in radius, a region inaccessible to either messenger alone.
	
	These estimates carry several uncertainties. Astrophysically, the LRD BH mass function, the companion masses and, above all, the per-nucleus inspiral rate are poorly known, and the latter spans five orders of magnitude between ordinary and dense-cluster nuclei. On the modelling side, we assume circular orbits and  gas at rest, include only the enclosed mass for the disc, and fit only the coalescence time and phase, so our resolvability estimates are optimistic; each interior model also carries its own caveats (Sec.~\ref{sec:models}). Within these limits, LISA has the potential to probe the interiors of LRDs if they are dense and host multi-year inspirals at ordinary galactic-nucleus rates, preferably with companions of a few tens of solar masses.
	
	\section{Acknowledgments}

	This work was supported by the Project No. PID2022-138963NB-I00 and PID2025-168215NB-I00, funded by the Spanish Ministry of Science,
	Innovation and Universities (MCIN/AEI/10.13039/
	501100011033) and by the Comunitat Autonòma de les Illes Balears through a ``Ajut per a projectes de recerca científica i tecnològica (PRD2025, PRD$2025\_099$)''.
	X.J.\ is supported by the Spanish Ministerio de Ciencia, Innovaci\'on y Universidades (Beatriz Galindo, BG22-00034) and cofinanced by UIB.
	During preparation of this work, the authors used Claude (Anthropic) to assist with code debugging and language editing. The authors reviewed, verified, and take full responsibility for all content in the manuscript.

\end{document}